# Chirped pulse spectrometer operating at 200 GHz

Francis Hindle[1*], Cédric Bray[1], Daniele Fontanari[1], Meriem Mouelhi[1], Arnaud Cuisset[1], Gaël Mouret[1], Robin Bocquet[1]

[1]Laboratoire de Physico-Chimie de l'Atmosphère, Université du Littoral Côte d'Opale, 189A Av. Maurice Schumann, 59140 Dunkerque, France

Abstract

The combination of electronic sources operating at high frequencies and modern microwave instrumentation has enabled the recent development of chirped-pulse spectrometers for the millimetre and THz bands. This type of instrument can operate at high resolution which is particularly suited to gas phase rotational spectroscopy. The construction of a chirped pulse spectrometer operating at 200 GHz is described in detail while attention is paid to the phase stability and the data accumulation over many cycles. Validation using carbonyl sulphide has allowed the detection limit of the instrument to be established as function of the accumulation. A large number of OCS transitions were identified using a 10 GHz chirped pulse and include the 6 most abundant isotopologues, the weakest line corresponding to the fundamental R(17) transition of $^{16}O^{13}C^{33}S$ with a line strength of 4.3 x $10^{-26}$ $cm^{-1}$/(molec.$cm^{-2}$). The linearity of the system response for different degrees of data accumulation and transition line strength was confirmed over 4 orders of magnitudes. A simple analysis of the time domain data was demonstrated to provide the line broadening coefficient without the need for conversion by a Fourier transform. Finally, the pulse duration is discussed and optimal values are given for both Doppler limited and collisional regimes.

1. Introduction

The use of the millimetre/Terahertz (mm/THz) band ( 0.1 to 10 THz ) for the analysis of gas phase systems is attractive due to a vast number of polar compounds which produce strong rotational spectra. In particular, at low pressure, the molecular linewidths approach the Doppler limit and the narrowness of the spectral features ensures that an excellent selectivity can be achieved even in complex mixtures [1],[2]. In order to exploit these features, considerable attention has been paid to the technological development of radiation sources able to operate in this regime. At present, only amplified multiplier chains (AMC) and photomixer (PM) sources are practicable when high-resolution spectra are required for the analysis and quantification of gas phase species [3], [4]. Although Quantum Cascade Lasers (QCL) hold great promise as powerful THz sources, their suitability for gas analysis is dependent on overcoming the difficulties to calibrate, stabilise, and tune the emission frequency [5]. The majority of instruments developed for ground based mm/THz studies of gases employ a standard absorption spectrometer configuration. In this case, the maximum sensitivity is obtained by measuring a weakly absorbing line in the presence of a relatively intense background. In other words, a very small fractional power variation must be measured. This tends to limit the achievable sensitivity. Nevertheless the utility of AMC and PM instruments have been demonstrated; for both spectroscopic and analytical applications [4], [6]. The principal shortcomings of these solutions are the relatively long measurement time required to record the high-resolution spectra, and the limited sensitivity.

In contrast to the absorption spectroscopy employed in the mm/THz band, microwave (MW) instruments typically record a signal originating from a molecular emission. Classical Fourier Transform MW (FTMW) spectrometers developed by Balle and Flygare in 1979 [7] operate on this principle. A radiation pulse stimulates the gas under analysis. When the frequency corresponds to the molecular line frequency a macroscopic polarisation of the sample is induced. A coherent emission is then produced by the sample, which is termed Free Inductive Decay (FID). The FID can be recorded after the end of the excitation pulse overcoming the difficulties of the absorption configuration. The intensity of the FID decays as the sample gradually dephases and returns to its unpolarised state. A new technique employing a MW Chirped-Pulse (CP) and coherent FID detection has recently been developed [8]. In comparison with cavity-enhanced FTMW methods often employed at MW frequencies, an acquisition time some 50 times faster is realised for equivalent instrument sensitivities. Although the cavity enhanced technique has the advantage of a lower MW power requirement, it is severely penalised by the time consuming mechanical scanning of the cavity across the measurement bandwidth. Powerful MW solid state or traveling wave tube amplifiers are now available enabling the development of faster techniques such as CP spectroscopy where no mechanical scanning is required. The strategy employed in this case is to record the phase coherent FID with the largest possible bandwidth. Unlike FTMW spectroscopy no cavity is used. So the bandwidth of a single measurement cycle is only dependent on the frequency extension of the CP itself and the bandwidths of the mixer and oscilloscope employed for the detection of the FID signal. This has the advantage of being a multiplex approach, by recording the time dependent FID and taking the Fast Fourier Transform (FFT), the contributions from numerous molecular lines can be measured simultaneously. The potential of this type of instrument lies in the possibility of efficiently accumulating the signal over a large number of measurement cycles. The use of modern ultra-rapid arbitrary waveform generators and high-speed oscilloscopes/digitisers are critical for the implementation of CP instrumentation. At MW frequencies, a bandwidth of 11 GHz has been demonstrated in a single step [9]. This type of instrument is starting to be applied in the microwave band. It is however used with a jet-cooled molecular beam to populate the lower rotational states of heavy molecules and to produce weakly bounded molecular complexes. The use of a jet-cooled molecular beam is not required for the mm range as the maximum line intensities for the majority of light compounds are located in mm/THz at room temperature. At room temperature, mm/THz gas phase spectroscopy is now able to resolve increasingly heavy compounds with increasingly congested spectra coming from rotational transitions with high rotational quantum number values. Typically, the low-temperature molecular beam FTMW measurements are required prior to the room temperature mm/THz study. Indeed, in order to analyse these dense spectra, initial molecular beam FTMW measurents are used to provide the assignment of the low rotational quantum number energy levels. Subsequently, the ground state molecular parameters, deduced from the fit of molecular beam FTMW spectra, are used to extrapolate the mm/THz spectra for the assignment of a large variety of rotational transitions and in particular the higher rotational quantum number values [10]. Figure 1 displays, in the 190 - 210 GHz frequency range, pure rotational transitions of 12 atmospheric compounds referenced in the HiTRAN database [11] and monitored in the troposphere and stratosphere towards the Atmospheric Chemistry Experience (ACE) mission [12]. HiTRAN frequencies and intensities are plotted in linear and logarithmic scales, respectively. This figure highlights the large variety of rotational transitions which may be probed in a 20 GHz section of the mm range. Benefiting from the rapid acquisition time, CP-mm spectroscopy allows to optimize the SNR of rotational lines by averaging a large amount of spectra. We will demonstrate than $10^{-26}$ cm$^{-1}$/(molecule.cm$^2$) may be considered as the lowest intensity level detectable in the configuration of the CP-mm spectrometer described in this study. In a rough approximation, taking into account this value as a reasonable limit, a detection threshold in part.per.million (ppm) may be estimated. Some trace gas atmospheric species such as $SO_2$, ClO, OCS, HOCl, $H_2O_2$ may be detected in the 190 - 210 GHz CP-mm frequency range at subppm level. Moreover, in Figure 1 , the majority of rotational transitions with intensities from $10^{-24}$ to $10^{-26}$ cm$^{-1}$/(molecule.cm$^2$) are predicted in the JPL [13] and HiTRAN spectroscopic databases but have not been experimentally observed.

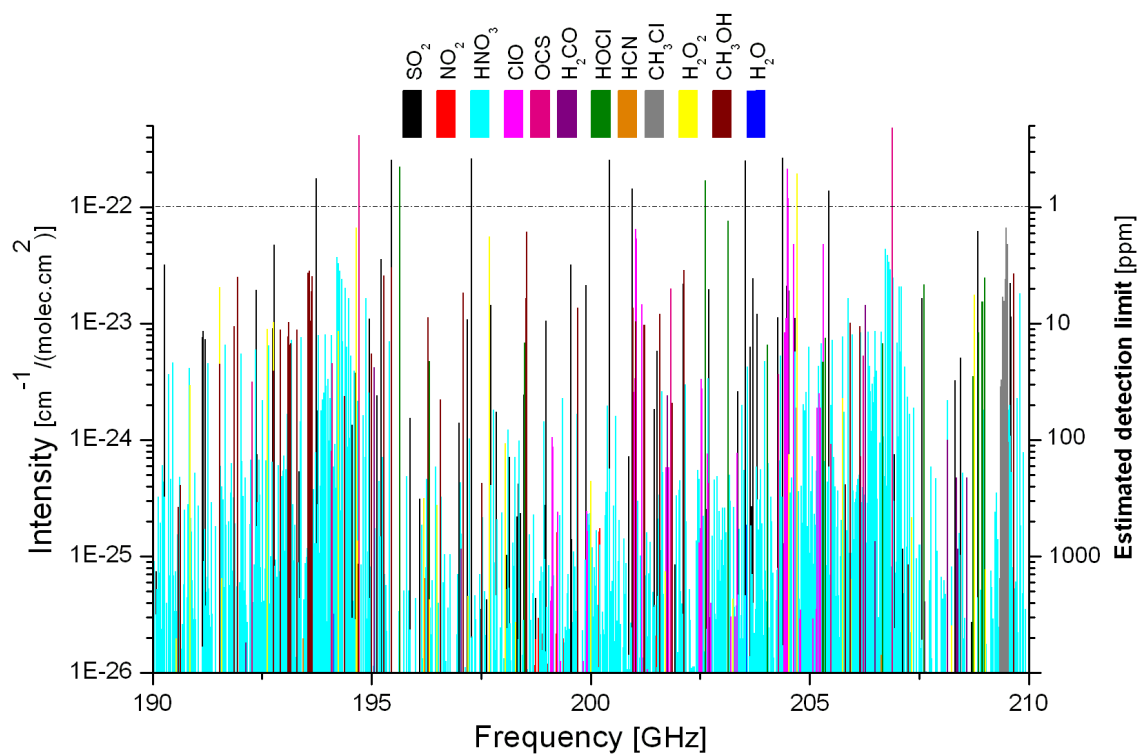

**Fig. 1: Rotational transitions of 12 atmospheric compounds, referenced in the HiTRAN database and monitored in the ACE mission, detectable by the CP-mm instrument. The HiTRAN line intensities (left) are plotted as a function of the line center frequencies. The second scale (right) provided indicates the estimated trace gas detection limit.**

Work undertaken at the University of Virginia has shown that the CP measurement modality developed in the MW can be transposed to mm/THz frequencies [14]. The CP may be generated at MW frequencies using the existing techniques and translated to the target frequency by an AMC. The first demonstration of a CP-mm instrument employed an AMC capable of delivering 30 mW at 275 GHz. It gave strong molecular signals of propionitrile ($CH_3CH_2CN$). One advantage of working in the mm band is that the frequency multiplication expands the CP bandwidth while conserving its duration. Unlike the MW band where a direct acquisition of FID can be undertaken using a fast oscilloscope or digitiser, no such instrumentation is available for mm frequencies. The FID must be down-converted using a heterodyne scheme to enable its measurement by standard MW frequency oscilloscopes, this also has the benefit of preserving the phase stability of the signal. The most convenient solution is a sub-harmonic mixer, a three port device producing an intermediate frequency as a function of its local oscillator and received signal. The intermediate frequency signal is a replica of the received signal, down-converted into the MW band where it can be digitised for further processing. The ability to efficiently combine the signals obtained from multiple CP cycles is key to the success of the technique. The complete CP measurement cycle consists of a CP excitation followed by a FID measurement period after complete extinction of the source. The duration of the FID signal is dependent on the relaxation time of the excited state of the probed rotational transition of the gas under analysis. A large number of measurement cycles are accumulated in the memory of the fast oscilloscope before taking the average of the temporal waveform and obtaining the spectrum by performing a FFT. So far, only a small number of CP instruments have been demonstrated in the mm/THz band [14]–[16]. Here we have constructed a CP instrument operating at 200 GHZ as a compromise between the available source power and the line intensities of numerous small polar molecules. Details of the spectrometer construction and validation with a stable well known molecule is used to determine the system performance. Numerous aspects will be discussed in the context of the optimisation of the instrument.

2. CP-mm spectrometer

The critical elements used to build a CP-mm instrument are an AMC and arbitrary waveform generator (AWG) for the pulse emission, while a sub-harmonic mixer and data acquisition system are used for the detection. The cycle-to-cycle phase stability and synchronisation of the chirped emission followed by a detection period are essential. They allow a large number of measurement cycles to be rapidly and efficiently accumulated. An AWG with two output channels using the same sampling points for the two channels is particularly useful in this regard. The first channel may be used to generate the chirp, while the second produces a monotone signal for the local oscillator of the sub-harmonic mixer, as illustrated in Fig 2.

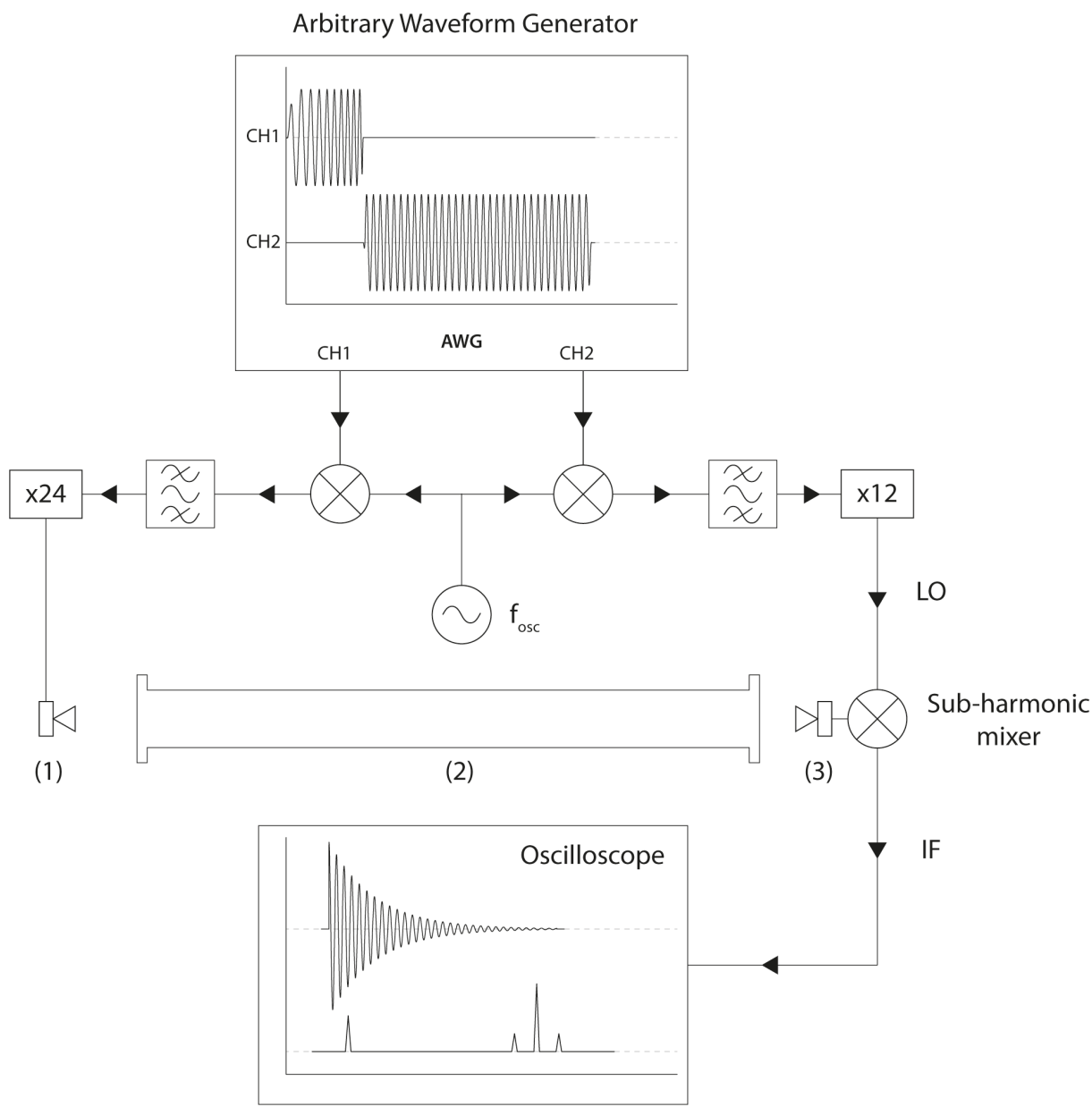

Fig 2. Chirped Pulse millimetre (CP-mm) spectrometer. The emitted CP is generated by mixing the AWG CH1 with a fixed oscillator ($f_{osc}$), before being frequency multiplied to the 190 to 210 GHz range. It is launched into free space using a diagonal horn antenna (1) and propagated through the sample chamber (2). A second diagonal horn antenna (3) is used to couple the molecular FID signal to the sub-harmonic mixer. The other ports of the mixer are used to provide the local oscillator (LO) signal and extract the intermediate frequency (IF). The IF signal is recorded using a fast oscilloscope whilst being triggered by a timing signal generated by the AWG.

In our case an AMC (Virginia Diodes) covering the band from 190 to 210 GHz, able to provide a power level of 30 mW, was employed. Operating with a multiplication rank of 24, microwave frequencies from 7.917 to 8.750 GHz must be supplied to obtain the required output. Arbitrary waveform generators able to directly generate these frequencies are available but remain particularly expensive. A lower specification AWG was used (Tektronix AWG 7122C) running at rates of up to 12 GSamples/s and with the capacity to generate a maximum frequency of 3.1 GHz. The AWG output was up-converted by mixing with the fixed frequency oscillator ($f_{osc}$), operating at around 12 GHz, in order to cover the required frequency range.

The mixer output contained the original carrier frequency $f_{osc}$, with the upper ($f_{osc} + f_{AWG}$) and lower ($f_{osc} - f_{AWG}$) sidebands. Only the lower sideband was required to generate the CP. It was isolated by a bandpass filter centred at 8.6 GHz, the upper sideband and the carrier frequency were heavily attenuated by the filter which presents a rejection of 30 dB in the stop bands. After filtering, the lower sideband signal was amplified to obtain 10 dBm at the input of the AMC. The chirped pulse generated in the microwave band was not only shifted to a higher frequency by the AMC but also benefits from a corresponding expansion of the chirp bandwidth. The optimal duration of the pulse is dependent on the targeted application, pulses in the range of 50 ns to 2 µs have been examined. The chirp stimulated the generation of the molecular FID signal to be detected as a function of frequency. This was undertaken by a heterodyne configuration to translate the FID emission down from around 200 GHz to the microwave band for measurement by an oscilloscope or data acquisition system. Indeed, the spectral width of the chirped polarisation and molecular FID should be conserved in this operation leading to a truly multiplex measurement over the targeted band. A sub-harmonic mixer was used (Virginia Diodes), the local oscillator (LO) frequency produced by a second AMC was feed to the first port of the mixer by a rectangular waveguide. Frequencies in the range of 95 to 105 GHz are required at the mixer and correspond to 7.917 to 8.750 GHz at the input of the LO AMC that has a multiplication rank of 12. The second channel of the AWG was used to generate this monotone signal in an identical manner to the baseband production of the chirped emission. The FID signal was coupled from free-space to the second port of the mixer using a horn antenna (WR-5.1) while the intermediate frequency (IF) signal was extracted from the third port via a standard coaxial connection. An oscilloscope with an analogue bandwidth of 12.5 GHz and maximum sampling rate of 100 GSamples/s (Tektronix DPO71254C) was used to measure the amplified intermediate frequency signal.

3. Example Signals

A typical chirped pulse duration of 250 ns was used and is followed by the acquisition of the IF signal for a period typically of 2 µs. The oscilloscope was used to collect and process the data from many measurement cycles. The time domain signal was averaged before taking the fast Fourier transform to obtain the FID signal spectrum. A Kaiser-Bessel apodization function was optimised to provide a suitable compromise narrow linewidths without introducing any secondary lobes. In order to maximise the throughput of the data processing the oscilloscope was specified with a memory of 250 MPts/channel. An example of the chirped pulse generated is given in figure 3. The frequency generated by the AWG $f_{AWG}$ was translated and multiplied to obtain the emitted frequency, $f_e = 24(f_{osc} - f_{AWG})$. The AWG output was defined by the carrier frequency and chirp width, 2.343 GHz and 41.7 MHz respectively. The emitted pulse was propagated through an empty measurement cell and coupled to the sub-harmonic mixer. An equivalent local oscillator frequency $f_{LO}$ of 197.28 GHz was used to down-convert the chirp corresponding to intermediate frequencies from 9 to 10 GHz, obtaining the desired chirp width of 1 GHz extending from 206.25 to 207.25 GHz.

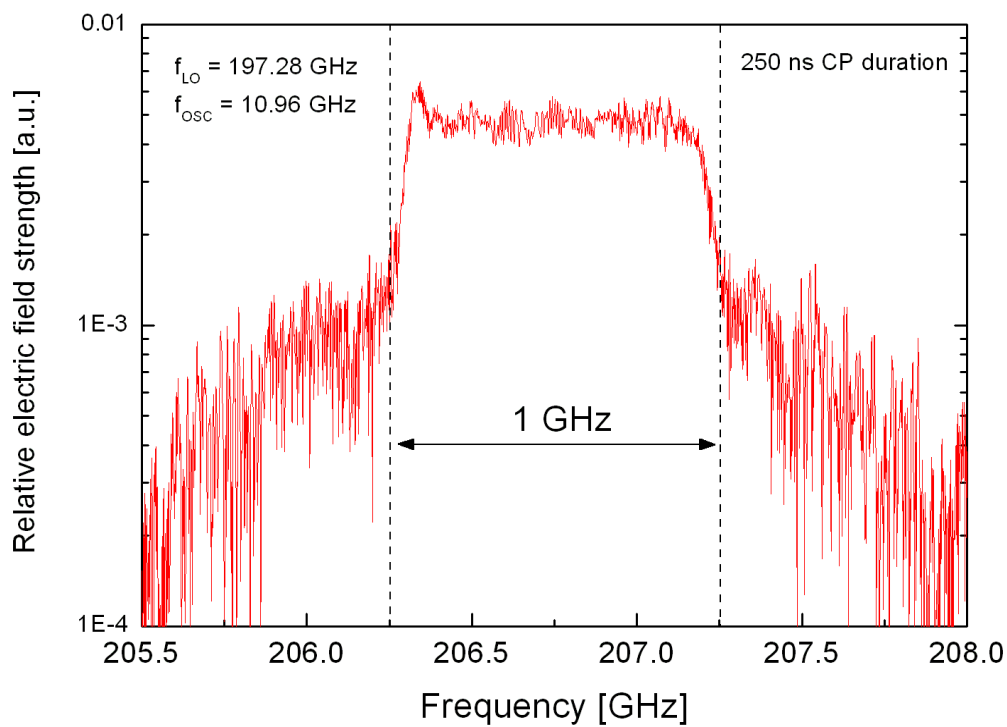

Fig 3. Emitted chirped pulse centred at 206.75 GHz with a width of 1 GHz, measured by heterodyne detection.

Typical free induction decay signals measured immediately after the extinction of the pulse are shown in figure 4 where the molecular responses of the Carbonyl Sulphide (OCS) linear-top and slightly asymmetric-top Methanol ($CH_3OH$), are compared. The envelope of the OCS FID signal displays a regular decay following the expected exponential function and is in agreement with equation 1 detailed in section 5. The accumulated oscillations can be visualised using a more detailed time scale as in the inset of figure 4. The response of methanol is somewhat weaker as is expected. However, the form of the envelope does not follow the same regular decay. This is due to the interaction of several closely spaced molecular lines, each contributing to the total FID signal. In the case of OCS a single strong line dominates the response whereas for methanol several weaker lines, all of similar strength, prevent the straightforward interpretation of the signal envelope.

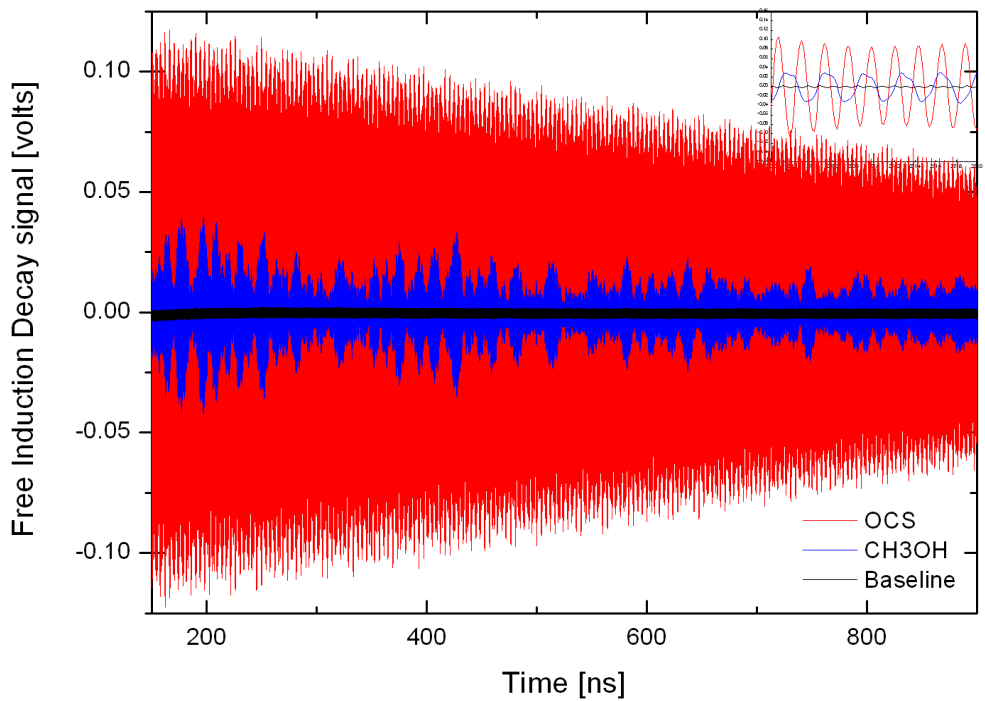

**Fig 4.** Free Induction Decay signal. The time domain measurements of Carbonyl Sulphide (Red trace) and Methanol (Blue trace). The inset figure shows a zoom covering the period from 250 ns to 252 ns. 1 000 spectra accumulation at a pressure of 5 µbars.

4. System performance

In order to evaluate the performance of the instrument, a series of validation spectra of OCS have been measured. OCS was chosen as its spectroscopy is well known and it presents a strong molecular $^{16}O^{12}C^{32}S$ ground state signature accompanied by a series of weaker transitions for the less abundant isotopologues and vibrationally excited states. Spectra for a chirp of 10 GHz have been taken with averaging ranging from a single shot acquisition n = 1 to n = $10^6$ measurement cycles. The spectrum for the highest degree of averaging is shown in figure 5, along with a baseline measurement and the expected line strengths. The sensitivity of the instrument leads to a large number of observed lines, a small number of peaks are caused by spurious electronic signals. The strongest spurious signal is the regeneration of the base band LO frequency by the sub-harmonic mixer. This effect was minimised by the empirical selection of the LO frequency which was finely adjusted. Nevertheless, several peaks remain. They are however easily identified and occupy a small fraction of the available frequency space (5 lines with a width of 8 MHz/line over a total of 10 GHz). The resolution of the spectrum is 1.9 MHz which results from the 400 ns gating of the Fourier transform (the optimisation of the resolution will be discussed in section 5). The assignment of the CP-mm OCS spectra has been performed using a previously published global rovibrational analysis [17]. The strongest molecular line at 206.745 GHz corresponds to the most abundant isotope $^{16}O^{12}C^{32}S$ and has an expected intensity of 5.8 x $10^{-22}$ cm$^{-1}$/(molec.cm$^{-2}$). The SNR obtained for this line defined as the ratio of the voltage amplitude to the background noise is evaluated to be in excess of $10^5$. The weakest tabulated isotope observed was $^{16}O^{13}C^{33}S$ having a line strength some four orders less intense. As expected, the noise level was observed to fall with the averaging as a function of $n^{-½}$. The measured noise levels are indicated in fig 5 for a single shot measurement, n = 100, and n = 10k. The SNR available for each of the six tabulated OCS isotopes are given in the inset as a function of the line strength and averaging. All the spectra demonstrate the desired linearity over several orders of magnitude. In the case of $n=10^6$, a dynamic range in excess of 4 orders of magnitude is available. The inset includes a linear fit, the deviations from the model are attributed to the response to a 350 ns pulse used to optimise the intensity of the signal. The compromise between the pulse length and uniformity of the instrument response to all frequencies excited by the chirp will be discussed in the following section. Other than the lines of the principal isotopologues in their ground state, the large majority (90%) of the observed transitions, have been attributed to vibrationally excited states populated at room temperature. For $^{16}O^{12}C^{32}S$, rotational transitions in 21 different excited states from $v_2 \leftarrow v_2$ (centred at 520.4 cm$^{-1}$) to $4v_2 \leftarrow 4v_2$ (centred at 2080.4 cm$^{-1}$) have been observed. The time overhead for processing the large volume of data is at present the limiting factor governing the maximum acquisition rate. For example, a single measurement cycle is composed of the CP, followed by a short wait time for the recovery of the amplifiers ( < 50 ns ), and the acquisition period corresponding to the time gated FFT. In this example for the measurement of OCS the cycle period was around 800 ns. If no time is lost for data processing or dead-time between cycles then, 1M cycles may be completed in under 1 s. Including the time overheads of the oscilloscope, the real elapsed time from the start of the first cycle to the end of the data processing for 1M cycles was approximately 10 minutes. For such very large datasets the oscilloscope memory is insufficient to hold the entire set and intermediate processing must be performed. The smaller datasets n = 1, n = 100, and n = 10k could be entirely stored in the oscilloscope memory and the elapsed start to finish times were 0.3 s, 1 s and 10 s respectively. If faster processing is required, a frequency segmented approach may be employed [16].

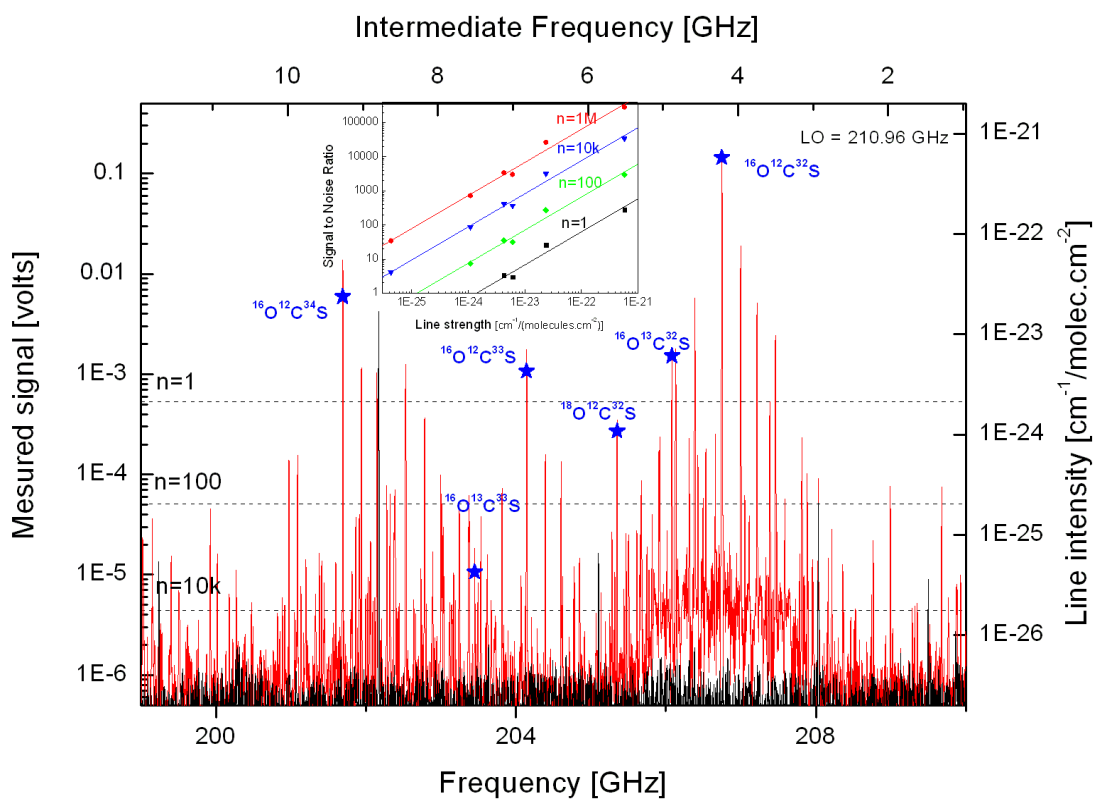

**Fig 5.** Measured spectra of OCS at 80 μbars using a chirped pulse centred at 205.008 GHz with a chirp width of 10 GHz and duration of 350 ns. The Fourier transform was gated to a period of 400 ns immediately after the pulse and recovery of the amplifiers. The red trace is the spectrum measured with accumulation of $10^6$ measurement cycles, black trace background signal under identical conditions. Blue stars show the expected ground state frequencies and line intensities [11], [17] for the 6 most abundant isotopologues of OCS. The noise levels for lower degrees of signal accumulation are also indicated (dashed lines). Inset, SNR of the molecular signal for each of the six isotopes taken over a varying number of measurement cycles. The measured data (points) are displayed with a linear fit (solid line).

5. Optimisation of the system parameters

The configuration and optimisation of such a CP-mm spectrometer is much less straightforward compared to instruments measuring absorption. Indeed, as the measurement cycle is divided into excitation and detection phases, if the excitation pulse period is increased strong variations of the response are observed for signals from the start and end of the pulse. However, if a degree of variation in the response can be tolerated, stronger signals may be obtained. Not only does the signal accumulation allow the dynamic range and measurement time to be controlled, but this kind of instrument can be adapted for a number of different applications by prioritising the optimisation criteria via the numerous system parameters. To identify the different measurement scenarios, the variation of signal amplitude as a function of CP duration and bandwidth should be understood. The observed FID signal is dependent on the CP and the molecular lineshape of the targeted transition. The first lineshape approximation is the well-known Voigt profile which represents the convolution of a Doppler broadened Gaussian profile, and a pressure broadened Lorentzian form. The corresponding signal amplitude of the FID is given by Φ (t) [18] :

$$\Phi(t) \propto \exp\left[-\left(2\pi\gamma P t + \frac{\pi^2}{\ln(2)}\Delta\nu_d^2 t^2\right)\right]$$

where,

$$\Delta\nu_d = \frac{f_0}{c}\sqrt{\frac{2\ln(2)k_B T}{m_a}}$$

γ represents the collisional broadening coefficient, $\Delta\nu_d$ the Doppler width (HWHM) with the Boltzmann constant $k_B$, the pressure P, the temperature T, the molecular mass $m_a$ of the absorber, c the speed of the light in vacuum and $f_0$ the center frequency of the transition. The envelope of the time domain FID signal can be used to directly determine the collisional broadening coefficient. In the case of the R(17) transition of OCS, a coefficient of 5.4±0.6 GHz.atm$^{-1}$ was obtained compared to 5.1±0.3 GHz.atm$^{-1}$ measured using a diode laser spectrometer in the $\nu_2$ band [19]. Examples of the fitted FID envelope are shown in Figure 6 for pressures of 75, 150 and 300 µbar that yielded time constants of 416, 208 and 104 ns respectively.

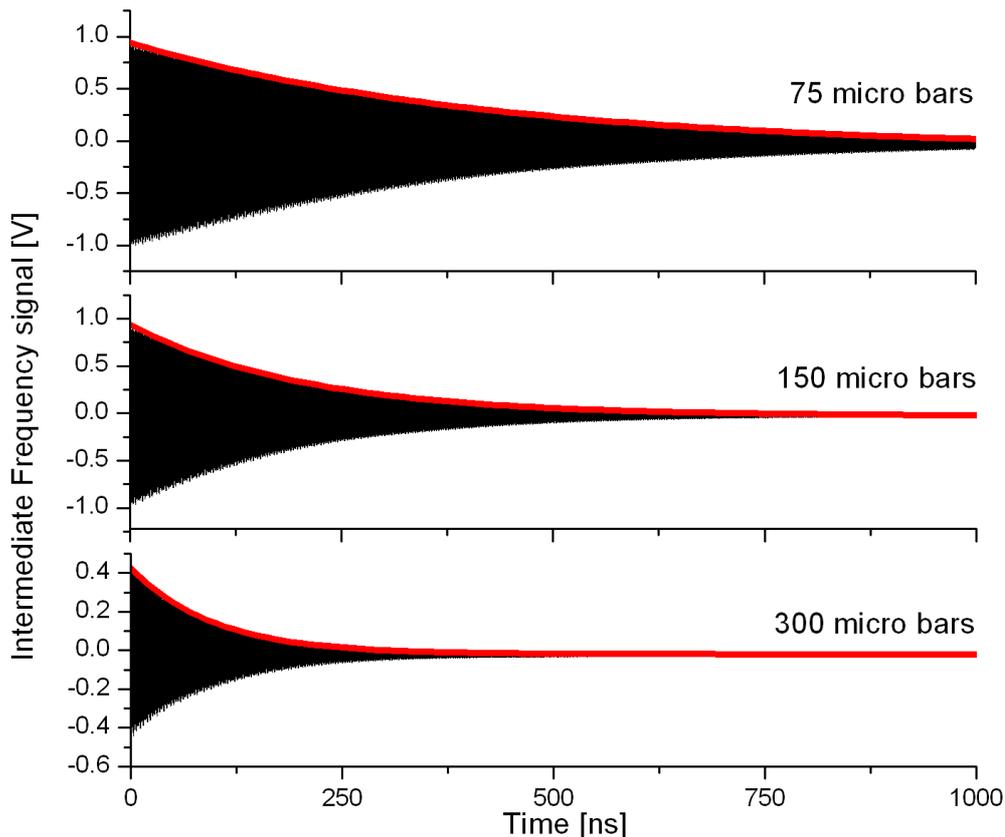

**Fig 6.** FID of OCS at pressures of 75, 150 and 300 µbar recorded for 1 µs. Red lines represent the fitted envelope of the measured FID signals. The R(17) transition of $^{16}O^{12}C^{32}S$ at 206.745 GHz, intermediate frequency 4.215 GHz, accumulation of 1000 measurement cycles.

The amplitude of the FID signal detected is strongly dependent on the polarisation created by the chirped pulse. For microwave spectroscopy, pulses with a linear frequency sweep have been shown [20] to be particularly efficient at polarizing molecular samples. The FID signal amplitude resulting from a chirped pulse displays the following dependencies :

$$S \propto \omega_0 E_{pulse} \mu^2 \Delta N \sqrt{\frac{\pi}{\alpha}}$$

where µ is the permanent dipole moment, $E_{pulse}$ the electric field strength, $\omega_o$ the line center frequency, and $\Delta N$ the population difference at equilibrium. It is assumed that $\Delta N$ is unchanged by the pulse and proportional to pressure. For chirped pulse excitation, $E_{pulse}$ is constant for all frequencies in the pulse and is fixed by the peak power of the amplifier, α is the sweep rate. Scaling of signal with $\alpha^{-\frac{1}{2}}$ is observed in Figure 7. At chirp bandwidths of 12.5 MHz and above the measured signal amplitude follows the expected relation with the sweep rate. For smaller bandwidths the response no longer follows this model and tends towards the response for a single frequency or monotone pulse, indicated by the blue asymptote.

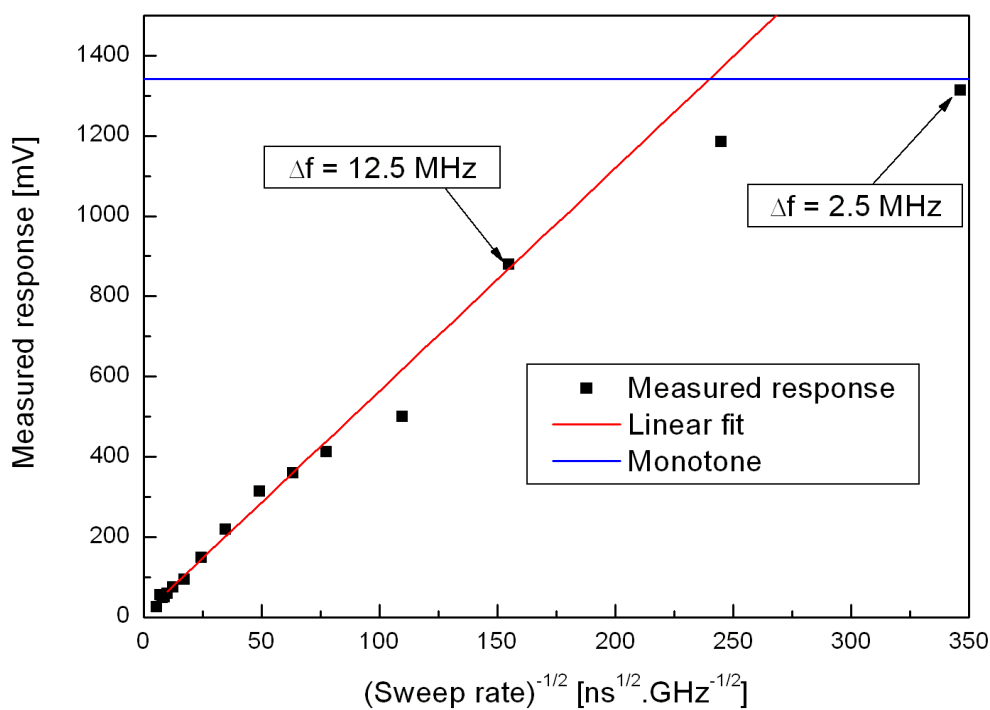

Fig 7. Measured amplitude of OCS transitions at 206.745 GHz with a chirp duration of 300 ns. Chirp bandwidth is varied from 2.5 MHz up to 10 GHz. The graph is linearized by plotting the amplitude as a function of sweep rate$^{-1/2}$. Red line is a linear fit of amplitude versus the sweep rate$^{-1/2}$. Blue horizontal line is the measured response when a monotone pulse at the line center frequency is used instead of a CP.

The measurement cycle of the CP-mm instrument is composed of two subsequent steps. Firstly, the polarisation of the molecular sample by the chirped pulse where the frequency of the emission is rapidly scanned across the measurement band exciting the corresponding transitions. The second phase, for the detection of the FID signals, is begun after the extinction of the source and once the detection amplifiers have recovered from being exposed to the powerful chirped pulse. The amplitude of any observed FID is naturally dependent on the delay between the time of observation and the initiation of the signal. This is not only of importance for the amplifier recovery time but for the initiation time difference between the lines at the start and end of the chirp. A compromise must be sought between the pulse duration and the uniformity of the instrument response to the molecular lines with center frequencies corresponding to the start and end of the CP. A short pulse will produce weak FID signals with an amplitude directly related to the line strength in question. As the pulses are lengthened the amplitude of the signals will increase, however the response of the lines whose FID was initiated at the start of the pulse will be weaker than those at the end of the pulse. This is a direct consequence of the additional delay from the start of the pulse to the acquisition period. A relatively simple model can be established to examine behaviour and assist in the optimisation. The FID initiation time is determined for all the frequencies in the CP. At a given pressure the relative spectrometer response is obtained by calculating the signal strength of each frequency when observed at the end of the pulse. In this manner, the additional variable delay encountered by the lines at the start of the pulse compared to those at the end is accounted for. The relative response has been determined as a function of the position of the transition and pulse length, as shown in figure 8. Here a pressure broadened regime is examined with a pressure of 90 µbar for pulse durations from 50 to 1000 ns. The characteristic relaxation time ($T_2$) for OCS at this pressure is 370 ns. The 50 ns pulse suffers from a weak average response but has almost no amplitude variation from start to finish. The variation of the response is clearly observed at 400 ns and confirmed by the experimental data. At 1000 ns the response to any lines positioned at the start of the pulse is almost zero whereas those at the end display to strongest response. The experimental data where measured using a single isolated transition while modifying the CP to place the transition at different positions.

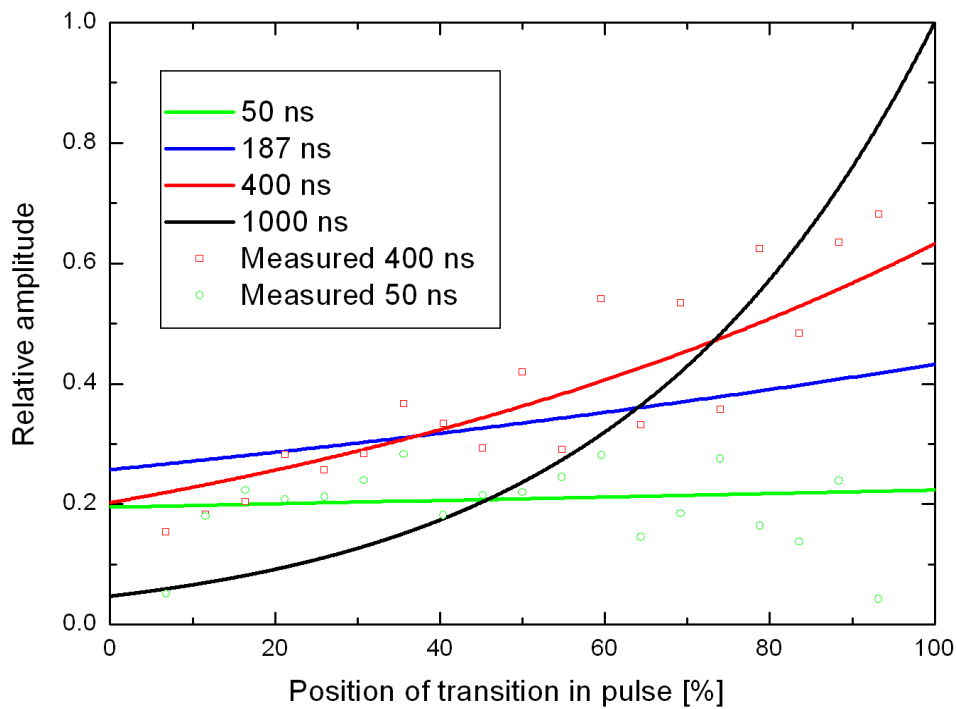

**Fig 8. Simulation of relative amplitude of a FID signal when observed at the end of the pulse as a function of the position of the transition in the chirped pulse (solid lines). Position is expressed as a percentage, 0 % corresponding to the start of the pulse and 100 % the end. Calculation realized for pressure broadened OCS (90 µbar). The simulations for 50 ns and 400 ns are compared to the measured experimental responses obtained by varying the CP carrier frequency.**

The observed variation is strongly dependent on the pressure which defines the relaxation characteristics in the time domain and linewidth for a frequency based representation. Indeed, we suggest an optimal duration pulse can be determined by maximising the response to lines at the start of the pulse if a degree of response variation can be tolerated. For collisional broadening, the optimum is given by:

$$T_{opt\ col} = \frac{1}{4\pi\gamma P}$$

whereas for Doppler limited lines dominated by the Gaussian component, providing that the time delay from the initiation of the FID to the start of the FFT is less than $1/\Delta\nu_d$, then the optimal pulse duration is:

$$T_{opt\ Dop} = \frac{\sqrt{\ln(2)}}{2\pi\Delta\nu_d}$$

The Fast Fourier Transform has a high impact on the optimisation of the chirped pulse experiment. In case of OCS at 200 GHz, the Doppler limited linewidth is 160 kHz (HWHM) requiring a measurement time window of 6.25 µs to reach an equivalent resolution. The FID signal decay time is around 1.6 µs, hence as the time window is increased to improve the resolution, the SNR decreases. This is demonstrated in figure 9 by the spectra of OCS at 8 µbars for time windows of 1 and 10 µs. The resolution offered by the shorter acquisition is limited while the SNR is in excess of three orders of magnitude. The longer window gives access to the expected Doppler limited profile with a SNR of 2 orders of magnitude. One strategy that can be adopted to optimise the utility of the data obtained in this manner is to employ zero padding. This results in larger number of frequency points but no additional information. The spectra for pressure broadened OCS (80 µbars) are shown in figure 9 for time windows of 1 and 10 µs. They are compared to an acquisition of 1 µs where zero padding has been applied to reach an equivalent of 10 µs. Although the visual aspect of the zero-padded spectrum is improved the SNR and resolution remain unchanged from the 1 µs spectrum. The 10 µs window shows the expected pressure broadened line which may be directly compared to the Doppler limited line measured at 8 µbars.

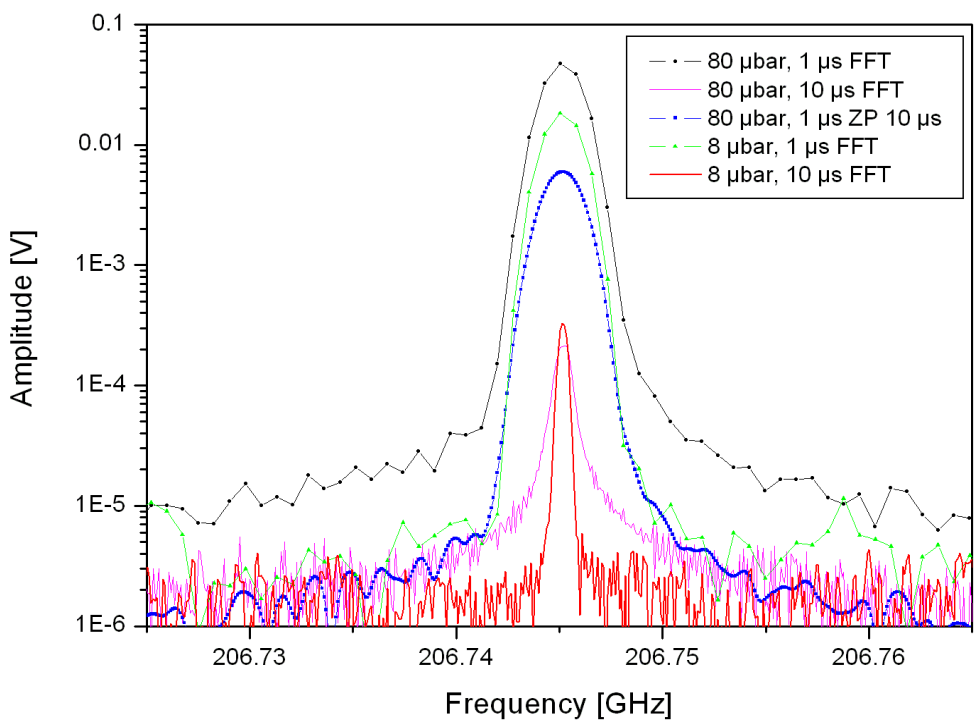

**Fig 9. Transition of OCS recorded at 8 and 80 μbar for Fourier transform windows between 1 or 10 μs and zero padding from 1 μs to 10 μs.**

The inevitable compromise between SNR and resolution should be addressed as a function of the requirements of the different applications. For example, highly congested spectra need the best resolution so the spectra should be taken with Doppler limited lines and measurement window sufficiently long to achieve a resolution equal to the Doppler width. At present the best resolution that can be achieved with the instrument is 100 kHz. In the case where the detection limit of a trace species is to be optimised the pressure broadened regime is more suitable. Indeed, spectral resolution can be sacrificed to give a stronger signal. A shorter measurement window is acceptable with the reduced decay time. Here a total real measurement time of 10 seconds was needed to recorded 1000 spectra with a time window of 10 μs, compared to 1 s for a time window of 1μs. Zero padding of the data can be used to improve the visual aspect of spectra. It does not however reduce the broadening of transition since no signal information is added.

6. Conclusion

The use of chirped pulses is now often employed in the microwave band with a few instruments operating in the mm and THz band. Here we have developed an instrument operating at around 200 GHz due to the availability/performance of the system components, and the large number of molecules which have a reasonably intense molecular signature at room temperature. Molecular lines with an intensity in the order of $10^{-23}$ cm$^{-1}$/(molec.cm$^{-2}$) can be recorded with a single measurement, a process taking approximately 1 μs which can easily be synchronised to an external signal if undertaking time dependent studies. The phase stability of the CP generation and detection system is demonstrated to provide an improvement in the SNR related to $n^{-\frac{1}{2}}$, which is accompanied by a linearity displayed over 4 orders of magnitude of line intensity. This type of instrument is of particular utility as the accumulation and sensitivity can be rapidly matched to the needs of many applications, such as low temperature reactivity studies where time resolved concentration monitoring of multiple species is required. The accumulation of $10^6$ measurement cycles allows transitions with intensities of $10^{-26}$ cm$^{-1}$/(molec.cm$^{-2}$) to be observed. One shortcoming of this approach is the time overhead of the required data processing, developments in dedicated high-speed acquisition and data treatment systems will without doubt provide new solutions to this limitation in the near future. The trade-offs between the needs of sensitivity, uniform spectral response and resolution are presented and optimal values for the pulse durations are given for different conditions.

7. Acknowledgements

The authors thank Marc Fourmentin (LPCA) for his assistance in the preparation of the figures for this manuscript. We would also like to acknowledge the French Agence Nationale de la Recherche for their financial support of this work via funding of the project Original Sub-millimeter Chirped pulse instrumentation for Astrochemical Reactivity (OSCAR) under contract number ANR-15-CE29-0017. The authors are also grateful for the financial support received from the Région Hauts-de-France, the Ministère de l'Enseignement Supérieur et de la Recherche (CPER Climibio), and the European Fund for Regional Economic Development through the TERAFOOD project (INTERREG V FR-WA-VL 1.2.11).